\documentclass[%
preprint,
 amsmath,amssymb,
 aps,
]{revtex4-2}

\usepackage{graphicx}
\usepackage{dcolumn}
\usepackage{bm}

\usepackage{makecell}
\usepackage{booktabs} 
\usepackage{siunitx}

\usepackage{comment}
\usepackage{amsfonts}

\begin{document}


\title{Logarithmic profiles of velocity in stably stratified atmospheric boundary layers}

\author{Yu Cheng}
 \email{yc2965@columbia.edu}
\affiliation{%
Department of Earth and Planetary Sciences, Harvard University, Cambridge, Massachusetts, USA\\
}%
\author{Andrey Grachev}%
\affiliation{%
 Boundary Layer Research Team/Atmospheric Dynamics \& Analytics Branch, DEVCOM Army Research Laboratory, WSMR, NM, USA\\
}%
\author{Chiel van Heerwaarden}%
\affiliation{%
	Meteorology and Air Quality Group, Wageningen University, Wageningen, the Netherlands\\
}%




\date{\today}

\begin{abstract}
The universal velocity log law first proposed by von K\'arm\'an in the near-wall region of turbulent shear flows is one of the cornerstones of turbulence theory. When buoyancy effects are important, the universal velocity log law is typically believed to break down according to
Monin-Obukhov similarity theory (MOST), which has been used in almost all global weather and climate models to describe the dependence of the mean velocity profiles on buoyancy in the atmospheric boundary layer. 
In contrast to MOST, we propose new logarithmic profiles of near-wall mean velocity in the stably stratified atmospheric boundary layers based on direct numerical simulations and field observations across a wide range of buoyancy effects. We find that buoyancy does not change the logarithmic nature of velocity profiles but instead modifies the slope of the log law in stably stratified conditions.
\end{abstract}

\maketitle


\section{Introduction}

In the near-wall region of turbulent shear flows, there exists a universal logarithmic velocity profile \citep{karman1930mechanische,millikan1938critical}, characterized by a constant slope between mean velocity and the logarithm of the distance to the wall.
The universal velocity log law has been supported by laboratory measurements of pipe flow \citep{mckeon2004further} and boundary layer \citep{monkewitz2007self}, atmospheric observations \citep{andreas2006evaluations}, and direct numerical simulations (DNS) \citep{lee2015direct,yamamoto2018numerical} of turbulent shear flows.
When turbulent flow is influenced by buoyancy, the mean velocity may not be adequately described by the universal log law. 
To address the buoyancy effects, Monin-Obukhov similarity theory (MOST) was proposed to revise the universal velocity log law using stability correction functions of the distance to the wall $z$ and the Obukhov length $L$ \citep{obukhov1946turbulence} based on dimensional analysis \citep{monin1954basic}.
MOST has been used to describe buoyancy corrections of the mean velocity profile and to provide velocity and momentum flux in the atmospheric surface layer, roughly the lowest $10 \%$ of the atmospheric boundary layer (ABL) \citep{stull1988introduction}, in almost all numerical weather prediction and climate models \citep{deardorff1972parameterization,troen1986simple,holtslag1993local,louis1979parametric}. According to MOST, the mean velocity profile is not logarithmic in stably stratified conditions as compared to von K\'arm\'an's universal log law due to buoyancy corrections.

The stably stratified conditions are frequently observed over land at night \citep{mahrt1998stratified} and in polar regions of the Earth \citep{grachev2015similarity} when the air is cooled by the land surface. 
Stably stratified turbulence in the ABL is very difficult to be represented \citep{viterbo1999representation,cuxart2006single,svensson2009analysis} in numerical weather prediction (NWP) and climate models \citep{holtslag2013stable,teixeira2008parameterization,svensson2011evaluation}.
The turbulence representation in the ABL is especially important for the Arctic as climate change is amplified there \citep{holland2003polar}.
The difficulty of representing stably stratified turbulence is partly due to the widely known failure of MOST in the atmospheric surface layer in very stable conditions \citep{mahrt1998stratified,mahrt2014stably}. 
In particular, MOST does not capture the mean velocity profile when buoyancy-driven stratification is significant as was shown in field observations of the very stable ABL \citep{forrer1997turbulence,pahlow2001monin,klipp2004flux,cheng2005pathology,grachev2005stable,yague2006influence}.

Based on a reformulation of MOST, Grachev et al. \cite{grachev2015similarity} proposed a similarity theory using the Dougherty-Ozmidov length scale $L_O$ \citep{dougherty1961anisotropy,ozmidov1965turbulent}, which is typically regarded as the outer scale of isotropic turbulence in stably stratified conditions \citep{gargett1984local,waite2011stratified,li2016connections,cheng2020model}. It is thus possible that some important length scales might be missing in the dimensional analysis of MOST when stably stratified turbulence is considered, as was shown in convective conditions \citep{tong2019multi,cheng2021logarithmic}.
The strength of stratification is typically described by the stability parameter $z/L$ according to MOST, where  $L=\frac{u_{\tau}^3}{\frac{\kappa g }{\Theta_r}u_\tau \theta_*}$, $u_\tau$ is the friction velocity, $\kappa$ is the von K\'arm\'an constant, $g$ is the gravitational acceleration, $\Theta_r$ is a reference potential temperature, $\theta_* \equiv \frac{\nu_\theta}{u_\tau} \frac{\partial \Theta}{\partial z}_{|z=0}$ is a scaling temperature, and $\nu_\theta$ is the thermal diffusivity. 
When buoyant stratification increases, $z/L$ increases. In addition, Grachev et al. \cite{grachev2015similarity} showed that $z/L_O$ may also characterize buoyancy effects, where $L_O$ is the Dougherty-Ozmidov length scale \citep{dougherty1961anisotropy,ozmidov1965turbulent}.

It is worth noting that the collapse of turbulence in stably stratified conditions can be indicated by the parameter $\frac{L}{\delta_v}=\frac{L u_\tau}{\nu}$ \citep{flores2011analysis}, where $\delta_v$ is the viscous length scale, and $\nu$ is the kinematic viscosity. $L/\delta_v$ can characterize both buoyancy effects (i.e., the inverse of gradient Richardson number) \citep{ansorge2014global} and Reynolds number effects (i.e., the scale separation between $L$ and $\delta_v$) \citep{flores2011analysis}.
We can write $\frac{L}{\delta_v}=\frac{z_i/\delta_v}{z_i/L}$, where $z_i$ is the boundary layer height. $\frac{z_i}{\delta_v} =\frac{u_\tau z_i}{\nu}$ can represent Reynolds number effects, and $\frac{z_i}{L}$ can represent buoyancy effects (e.g., in the convective boundary layer \citep{cheng2021logarithmic}). We will show that $\frac{z_i}{\delta_v}$ and $\frac{z_i}{L}$ can be used to constrain the slope of our proposed velocity log law.

Recently, the logarithmic temperature profiles have been reported in the near-wall regions of turbulent Rayleigh-Bénard convection \citep{ahlers2012logarithmic,grossmann2012logarithmic,ahlers2014logarithmic}, vertical natural convection \citep{holling2005asymptotic}, and convective ABL \citep{cheng2021logarithmic}, which is in contrast with the breakdown of a log law according to MOST. Thus, the logarithmic nature does not necessarily break down under the influence of buoyancy. In this study, we aim to investigate the existence of logarithmic velocity profiles in the stably stratified ABL and the possible dependence of velocity profiles on other stability parameters (or length scales) using high-resolution DNS experiments and field observations. 

\section{methods}
\subsection{Direct numerical simulations} \label{loglawSec2}

Large eddy simulations (LESs) \citep{moeng1984large,deardorff1972numerical,nieuwstadt1993large} have been widely used to study the ABL. However, subgrid-scale turbulence models  \citep{li2018implications,mellado2018dns} may lead to uncertainties near the wall and LESs might have difficulties in simulating strongly stratified turbulence \citep{jimenez2005large,flores2011analysis}. In addition, wall-modeled LES for the ABL usually invokes MOST \citep{moeng1984large,schmidt1989coherent,khanna1997analysis,bou2005scale,cheng2017failure}.
Moreover, turbulence spectra are often not well resolved \citep{cheng2020model} since
the Dougherty-Ozmidov scale is typically not resolved in LESs \citep{beare2006intercomparison,khani2014buoyancy,waite2011stratified} except possibly in a few studies \citep{sullivan2016turbulent}.
Recently, DNS has been used to study the stably stratified ABL \citep{flores2011analysis,ansorge2014global,shah2014direct,gohari2017direct,cheng2020model}, although the Reynolds number is not as high as that in the real ABL.
To obtain high-resolution velocity profiles in the near-wall region, DNSs of the stably stratified Ekman layers are conducted in this study.

The incompressible Navier-Stokes equations with Boussinesq approximation are solved \citep{heerwaarden2017microhh}. Periodic boundary conditions are employed in the horizontal ($x$ and $y$) directions. 
Firstly, we simulate a turbulent Ekman layer flow \citep{coleman1992direct} over a smooth surface in the absence of buoyancy as in previous studies \citep{shah2014direct,gohari2017direct}.
The three simulations of neutral Ekman layer flow named ReD900, ReD1800 and ReD2700 are forced with varying mean geostrophic wind. The grid points are $320 \times 320 \times 1664$ for the dataset ReD900, $640 \times 640 \times 3328$ for the dataset ReD1800, $960 \times 960 \times 4992$ for the dataset ReD2700 in streamwise ($x$), spanwise ($y$), and vertical directions ($z$), respectively.
The Reynolds number is $Re_D=\frac{U_g D}{\nu}$, where $U_g$ is the geostrophic wind speed, $D={\big( 2\nu/f \big)}^{1/2}$ is the laminar Ekman layer depth, $\nu$ is the kinematic viscosity, and $f$ is the Coriolis parameter.
Similarly to previous experiments \citep{shah2014direct,gohari2017direct}, a neutral velocity log law of the Ekman layer is obtained after $ft=5.9$, $6.0$ and $13.6$ for ReD900, ReD1800 and ReD2700, respectively. 
Then we add a cooling surface buoyancy flux $B_0$ to generate various stably stratified conditions. The boundary conditions for the temperature field are zero heat flux at the top of the computational domain. At the top $25\%$ of the computation domain, a sponge layer is added to prevent reflection of gravity waves \citep{nieuwstadt1993large}. 
The near-surface stability is measured by normalized Obukhov length $L^+=\frac{L}{\delta_v}=\frac{u_{\tau}^3}{\frac{\kappa g }{\Theta_r}u_\tau \theta_*} \frac{u_\tau}{\nu}$.
The initial $L^+(t=0)$ is used to measure the strength of imposed stratification, which is computed from $u_\tau$ in the neutral Ekman layer before the cooling surface buoyancy flux $B_0$ is applied. 
Details of the DNS setup can be found in Cheng et al. \cite{cheng2020model} and the code is described in Heerwaarden et al. \cite{heerwaarden2017microhh}.

We note that turbulence decays fast in the stably stratified Ekman layer \citep{gohari2017direct}, thus we only analyze the periods when $\frac{L_O}{\eta} >1 $ for the possible existence of Kolmogorov's energy cascade \citep{cheng2020model}, where $\eta=\big(\nu^3/\epsilon \big)^{1/4}$ is the Kolmogorov scale \citep{kolmogorov1941} and $\epsilon$ is the turbulent kinetic energy (TKE) dissipation rate. 
The friction Reynolds number $Re_\tau=u_\tau \delta_t/\nu$ at the selected time step of the DNS experiments ReD900 ($L^+=160$), ReD1800 ($L^+=800$), ReD1800 ($L^+=3200$), and ReD2700 ($L^+=160$) are 861, 1208, 1026 and 3122, respectively, where $\delta_t=u_\tau/f$ is the turbulent Ekman layer length scale. Following the suggestion of Shah and Bou-Zeid \citep{shah2014direct}, we compute the boundary layer height $z_i$ in DNS experiments as the height where maximum of velocity occurs. The eight stably stratified DNS experiments are described in Table \ref{tab:loglawdataDNS}.

\begin{table}
	\caption[Details of the DNS set-up.]{Key parameters of the simulated stably stratified ABLs. $Re_\tau=\frac{u_\tau \delta_t}{\nu}$ is the friction Reynolds number, $u_\tau$ is the friction velocity, $\delta_t$ is the turbulent Ekman layer length scale, $\nu$ is the kinematic viscosity, $z_i$ is the boundary layer height determined from the height where maximum of velocity occurs, $L$ is the Obukhov length, $L_x$, $L_y$ and $L_z$ are the domain sizes in the $x$, $y$ and $z$ directions, respectively. $\Delta_x^+=(\Delta_x u_*)/\nu$, $\Delta_y^+$ and $\Delta_z^+$ are the spatial grid resolutions denoted by inner units in the $x$, $y$ and $z$ directions, respectively. $\kappa_{u}$ is the inverse of the velocity log law slope. The range of velocity log law is also indicated using $z^+$ and $\frac{z}{L}$.}
	\begin{center}
		\def~{\hphantom{0}}
		\begin{tabular}{lccccccccc}
			\hline \hline
			\makecell{DNS\\data}  &   \makecell{$Re_D$}     &  \makecell{$Re_\tau=\frac{u_\tau \delta_t}{\nu}$}  &  \makecell{$\frac{z_i}{L}$} &   \makecell{$\Delta_x^+$ \\ ($\Delta_y^+$)}  &
			$\Delta_z^+$  &
			\makecell{$L^+$}  &
			$\kappa_{u}$ & \makecell{Log-law\\range in \\ $z^+$} &\makecell{Log-law\\range in \\ $\frac{z}{L}$}  \\  [3pt]
			\midrule
			\makecell{ReD900}   & 900        & 861   & 5.4    & $6.5$    &$0.82$ &160   &0.33   &$81\sim100$ &$1.01\sim1.25$ \\	
			\midrule
			\makecell{ReD900}   & 900        & 806   & 2.2      & $6.3$  & $0.80$   &480  &0.28   & $59\sim70$ &$0.30\sim0.35$ \\	
			\midrule
			\makecell{ReD900}   & 900         & 548   & 0.9    & $5.2$  & $0.66$   &1600   &0.28  &$83\sim94$ &$0.26\sim0.30$ \\	
			\midrule   
			\makecell{ReD1800}   & 1800         & 1208   & 6.6    & $3.8$  & $0.49$  &800   &0.20  &$81\sim97$ &$0.84\sim1.01$ \\
            \midrule			
			\makecell{ReD1800}  & 1800 & 1088  & 2.0  & $3.6$   & $0.46$  &1600  &0.17  &$86\sim103$ &\makecell{$0.55\sim0.66$} \\	
            \midrule			
			\makecell{ReD1800}  & 1800 & 1026  & 3.7  & $3.5$   & $0.45$  &3200  &0.16  &$112\sim131$ &\makecell{$0.40\sim0.47$} \\			
            \midrule			
			\makecell{ReD2700}  & 2700 & 3122  & 42.0   & $4.1$   & $0.52$  &160  &0.24  &$100\sim121$ &\makecell{$3.04\sim3.67$} \\			
            \midrule			
			\makecell{ReD2700}  & 2700 & 1624  & 11.2  & $3.0$   & $0.38$  &1600  &0.15  &$102\sim122$ &\makecell{$1.16\sim1.38$} \\			
			\hline \hline	 
		\end{tabular}
		
		\label{tab:loglawdataDNS}
	\end{center}
\end{table}

\subsection{Field observations}
A tower of 213 m has been installed in the Cabauw Experimental Site for Atmospheric Research (CESAR) \citep{apituley2008overview,bosveld2020fifty}
(4.926\si{\degree} E, 51.97\si{\degree} N) in the Netherlands, where multi-level turbulence observations at 10 m, 20 m, 40 m, 80 m, 140 m and 200 m above a grass field are available in the ABL. 
%
We download a number of 30-minute data segments between 1:00 and 5:00 UTC in July 2019 from the CESAR data archive as the raw data, which has been quality controlled \citep{bosveld2020fifty}. These include velocity measurements from cup-anemometers at multiple levels and surface flux measurements from sonic anemometers at 3 m.
Through detecting the top of an elevated aerosol layer, the boundary layer height $z_i$ is measured by the Lufft CHM 15k ceilometer \cite{choma2019comparative}. We calculate the average of ABL height over each 30-minute segment as the raw data.

The raw data satisfying the following two conditions are further used to characterize the velocity profile: (1) the mean surface heat flux in the 30-minute sampling period has to be negative (i.e., heat transferred from air to ground); and (2) the boundary layer height $z_i$ is larger than 800 m. The first condition is used to select stably stratified ABL. The second condition is used to ensure that we include as many measurements as possible (especially those at 200 m) within the atmospheric surface layer through prescribing large boundary layer height $z_i$.
These two conditions lead to 40 different stably stratified 30-minute periods.

\section{results}
\subsection{Existence of a velocity log law}
\begin{figure}
	\centering
	\includegraphics[width=16cm,clip=true, trim = 18mm 5mm 0mm 00mm]{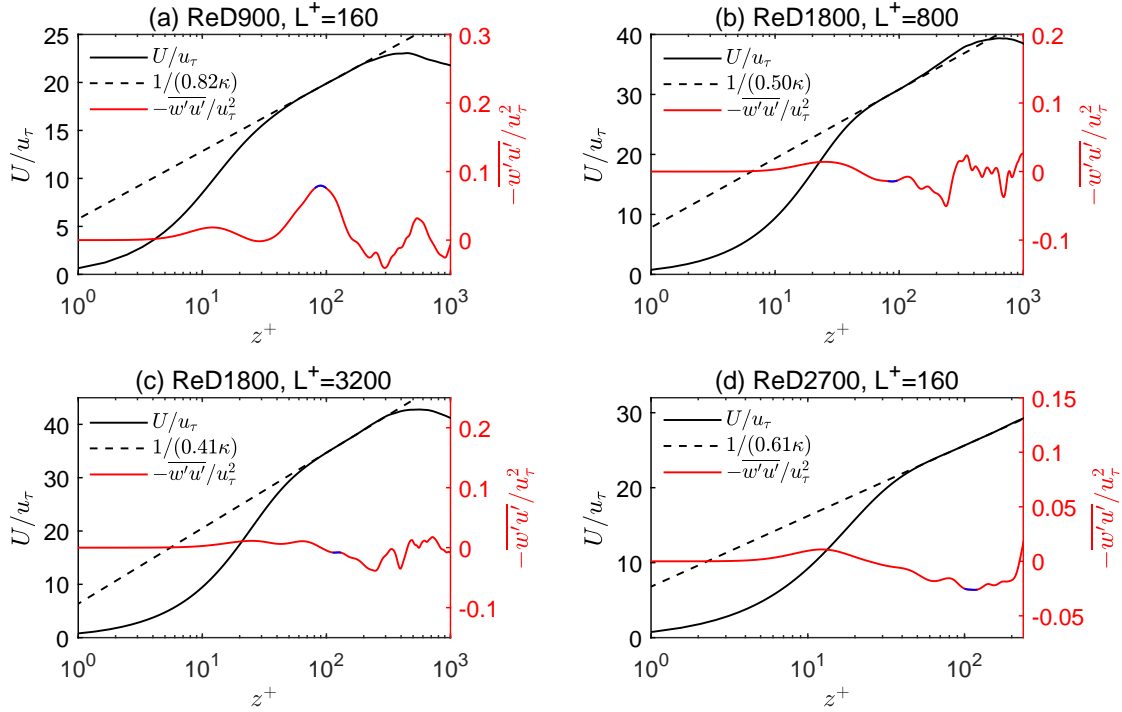}
	\caption{Normalized velocity $U/u_{\tau}$ and momentum flux $-\overline{w'u'}/u_{\tau}^2$ in the vertical direction of the DNS experiments. The blue line denotes the normalized momentum flux $-\overline{w'u'}/u_{\tau}^2$ in the velocity log law region.}
	
	\label{fig:loglawFig1}
\end{figure}

The normalized mean velocity $\frac{U}{u_\tau}$ fits a log law with $z^+ \equiv \frac{z}{\delta_v}$ in the DNS datasets (Fig. \ref{fig:loglawFig1}).
The coefficient of determination $R^2$ for $\frac{U}{u_\tau}$ and $\log(z^+)$ is 1.00 for the selected vertical layer (according to Fig. \ref{fig:DNSloglawKappaU}) near the wall across the DNS datasets.
Following Lee and Moser (2015) \cite{lee2015direct}, we use a plateau of $\frac{z}{u_\tau}  \frac{\partial U}{\partial z}$ to more clearly indicate the existence of a velocity log law (Fig. \ref{fig:DNSloglawKappaU}). The black dashed line (plateau) in Fig. \ref{fig:DNSloglawKappaU} is used to characterize the vertical layer where the velocity log law is identified (as detailed in Table 1). 

In the identified log law layer, the variations of turbulent momentum flux $\overline{w'u'}$ defined as $\frac{\max(\overline{w'u'})-\min(\overline{w'u'})}{\max(\overline{w'u'})}$ are $4.7 \%$ (ReD900, $L^+$=160), $10.6 \%$ (ReD1800, $L^+$=800), $5.3 \%$ (ReD1800, $L^+$=3200), and $4.3 \%$ (ReD2700, $L^+$=160), respectively. This is consistent with the definition of constant-flux layer \citep{stull1988introduction}, where flux variations are on the order of $~10\%$. This constant momentum flux zone is similar to the atmospheric surface layer, where the variation of turbulent
fluxes should scale with the ratio of the atmospheric surface layer
height to the ABL height \citep{wyngaard2010turbulence}, roughly $~10\%$.
The coexistence of a velocity log law and constant momentum flux in the DNS datasets (Fig. \ref{fig:loglawFig1}) resembles that in turbulent shear flows \citep{george2007there}.

As shown in Table 1, the range of Reynolds number of the DNS experiments is $548 \le Re_\tau \le 3122$, or equivalently $276 \le z_i/\delta_v \le 1345$. The vertical layers of velocity log law vary in different DNS experiments but all fall in the range $59 \le z^+ \le 122$. According to Marusic et al. \cite{marusic2013logarithmic}, the universal velocity log law of turbulent shear flows in the absence of buoyancy effects is found in the range $3 \left(z_i/\delta_v\right)^{1/2} < z^+ < 0.15 z_i/\delta_v $, which corresponds to $50 \le z^+ \le 202$ in our DNS experiments. Thus the range of the stably stratified velocity log law roughly falls within that of neutral velocity log law, although the influence of buoyancy effects cannot be neglected in our DNS experiments. In addition, the stably stratified velocity log law is found when $\frac{z}{L} > 3$ (in ReD2700, $L^+$=160) using the MOST stability parameter. However, MOST suggests that velocity profiles will significantly deviate from a log law in such stably stratified conditions (Fig. \ref{fig:DNSloglawKappaU}d). In fact, the stability correction functions of MOST are only defined in the range $0< \frac{z}{L} <1$ in stably stratified conditions due to its poor behaviour in more stratified conditions \citep{foken200650}. Therefore, the proposed velocity log law is fundamentally different from MOST and can be applied to a wider range of buoyant conditions.

The slopes of the proposed velocity log law are $\frac{1}{0.82 \kappa}$ (ReD900, $L^+$=160), $\frac{1}{0.50 \kappa} $ (ReD1800, $L^+$=800), $\frac{1}{0.41 \kappa} $ (ReD1800, $L^+$=3200), and $\frac{1}{0.61 \kappa} $ (ReD2700, $L^+$=160), respectively (Fig. \ref{fig:loglawFig1}). In comparison, the slope of the universal
log law for mean velocity in turbulent shear flows is constant, i.e., $\frac{1}{ \kappa} $ \citep{marusic2013logarithmic}. The variations of the slope of proposed velocity log law is due to buoyancy effects as well as Reynolds number effects. Similar dependence of the slope on buoyancy has been found in temperature log laws in the convective boundary layers \citep{cheng2021logarithmic} and Rayleigh-Bénard convection \citep{ahlers2012logarithmic}. The slope of the proposed velocity log law will be revisited later.

\begin{figure}
	\centering
	\includegraphics[width=16cm,clip=true, trim = 9mm 5mm 14mm 00mm]{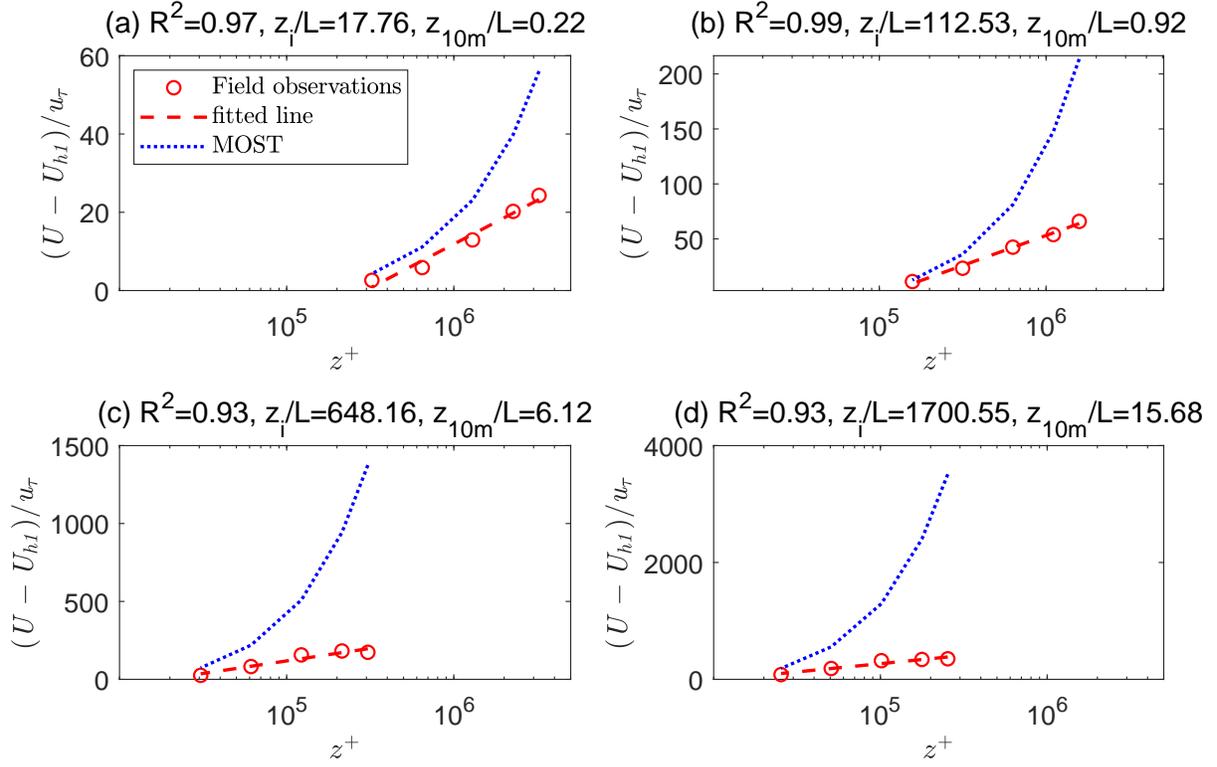}
	\caption{Normalized velocity $\frac{U-U_{h1}}{u_\tau}$ in the vertical direction of Cabauw observations. $R^2$ denotes the coefficient of determination for $\frac{U-U_{h1}}{u_\tau}$ and $\log(z^+)$, and ${z_{10m}/L}$ denotes the stability parameter $z/L$ at the height 10 m. ``MOST'' denotes the computed velocity profiles based on Monin-Obukhov similarity theory.}
	
	\label{fig:SHEBA_loglaw}
\end{figure}

In addition to the DNS experiments, we analyze field observations of the stably stratified ABL in the Cabauw experiment, which are at higher Reynold number ($1.4 \times 10^6 \le \frac{z_i}{\delta_v} \le 2.1 \times 10^7$).
A linear relation is fitted between the normalized velocity $\frac{U-U_{h1}}{u_\tau}$ and $\log(z^+)$ in the 4 sampled periods (Fig. \ref{fig:SHEBA_loglaw}), where $U$ is the temporally averaged wind speed at heights of 20 m, 40 m, 80 m, 140 m, and 200 m in a 30-minute period, and $U_{h1}$ is the averaged wind speed at 10 m. The coefficient of determination for $\frac{U-U_{h1}}{u_\tau}$ and $\log(z^+)$ satisfies the condition that $R^2>0.88$ in all the selected 40 periods, indicating the linear relation between $\frac{U-U_{h1}}{u_\tau}$ and $\log(z^+)$ and thus the presence of a velocity log law. 
We also compare the velocity profile based on MOST \citep{panofsky1963determination,kramm2013hans} with field observations and find substantial deviations across stably stratified conditions in the range $0.22 \le z/L \le 15.68$ (Fig. \ref{fig:SHEBA_loglaw}), where $z=10$ m (at one level of tower observations). 
The field observations in the real ABL confirm the existence of the velocity log law observed in our DNS experiments. 

It is worth noting that few studies have investigated the logarithmic nature of the mean velocity profile in the stably stratified ABL. This is partly due to the sparse measurements in the vertical direction. 
In addition, large uncertainties remain in the observations of the stable ABL since instruments often operate near their threshold levels due to small turbulence intensities \citep{nieuwstadt1984turbulent,mahrt1985vertical,mahrt2006extremely,mahrt2011near}.
Unlike DNS experiments, a plateau of $\frac{ z}{u_\tau}  \frac{\partial U}{\partial z}$ can hardly be identified in field observations to support a velocity log law. Moreover, field observations are often analyzed within the framework of MOST since MOST is still regarded as the foundation of ABL turbulence theory \citep{foken200650}.

 \subsection{Slope of the velocity log law}

\subsubsection{Dimensional analysis}

Fully developed stably stratified boundary layer flow can be described by $\nu$, $u_{\tau}$, $z$, $\theta_*$ and the boundary layer height $z_i$. 
 These variables can form 3 non-dimensional groups:
 \begin{equation}
 \frac{z}{L}=\frac{\kappa g \theta_*}{\Theta_r} \frac{z}{u_\tau^2},
 \end{equation}
 
\begin{equation}
 \frac{z_i}{L}=\frac{\kappa g \theta_*}{\Theta_r} \frac{z_i}{u_\tau^2},
\end{equation} 
%
and
\begin{equation}
\frac{z_i}{\delta_v}=  \frac{u_\tau z_i}{\nu}.
\end{equation} 
%
%
%
%
The mean velocity profile can be written as
 \begin{equation}
U=u_{\tau}F_0 \bigg(\frac{z}{L}, \frac{z_i}{L}, \frac{z_i}{\delta_v}  \bigg),
\end{equation}
where $F_0$ is a function of $\frac{z}{L}$, $\frac{z_i}{L}$, and $\frac{z_i}{\delta_v} $.
Following the argument for velocity gradient in Pope (2000) \cite{pope2000turbulent}, $\frac{\partial U}{\partial z}$ can be written as 
\begin{equation}
\frac{\partial U}{\partial z}=\frac{u_{\tau}}{z} \Phi \bigg(\frac{z}{L},\frac{z_i}{L},\frac{z_i}{\delta_v}  \bigg).
\end{equation}
According to the DNS datasets, $\frac{z}{u_\tau}  \frac{\partial U}{\partial z}$ is independent of $z$ in the log law region.
For the existence of a velocity log law, $\Phi$ has to be independent of $\frac{z}{L}$, leading to 
\begin{equation}
\frac{\partial U}{\partial z}=\frac{u_{\tau}}{z} \Phi \bigg(\frac{z_i}{L},\frac{z_i}{\delta_v} \bigg).
\end{equation}
We denote $\frac{1}{\kappa_u} \equiv \Phi \Big(\frac{z_i}{L},\frac{z_i}{\delta_v}\Big)$
and obtain
\begin{equation}
\frac{\partial U}{\partial z}=\frac{u_\tau}{z} \frac{1}{\kappa_u}.
\end{equation}
After integration from a reference height $z_r$ to $z$, we have 
\begin{equation}  \label{eqnLogprofile}
\frac{U-U_{z_r}}{u_\tau}= \frac{1}{\kappa_u} \log \Big( \frac{z}{z_r}\Big).
\end{equation}
Equation (\ref{eqnLogprofile}) is just the velocity log law since $\kappa_{u}$ is independent of $z$ and is a function of $\frac{z_i}{L}$ and $\frac{z_i}{\delta_v}$.
This dimensional analysis points out the relevant parameters that determine the slope of the proposed velocity log law, which should be calibrated from numerical experiments or field observations.

\begin{figure}
	\centering
	\includegraphics[width=16cm,clip=true, trim = 18mm 05mm 9mm 00mm]{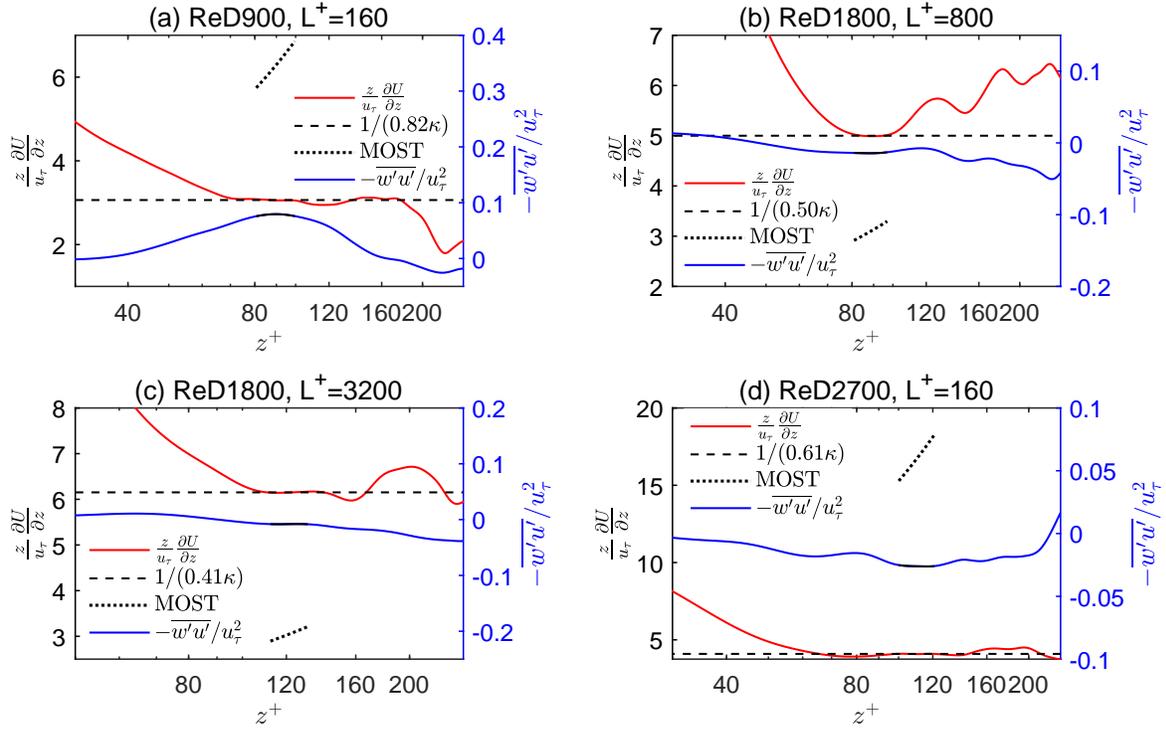}
	\caption{Vertical profiles of normalized velocity gradient $(z/u_\tau{\partial u}/{\partial z})$ (equal to $1/\kappa_{u}$) in DNS datasets. The widely used Monin-Obukhov similarity function of Businger et al. \cite{businger1971flux} denoted by ``MOST'' is also shown.}
	
	\label{fig:DNSloglawKappaU}
\end{figure}

 \begin{figure}
	\centering
	\includegraphics[width=16cm,clip=true, trim = 6mm 1mm 8mm 2mm]{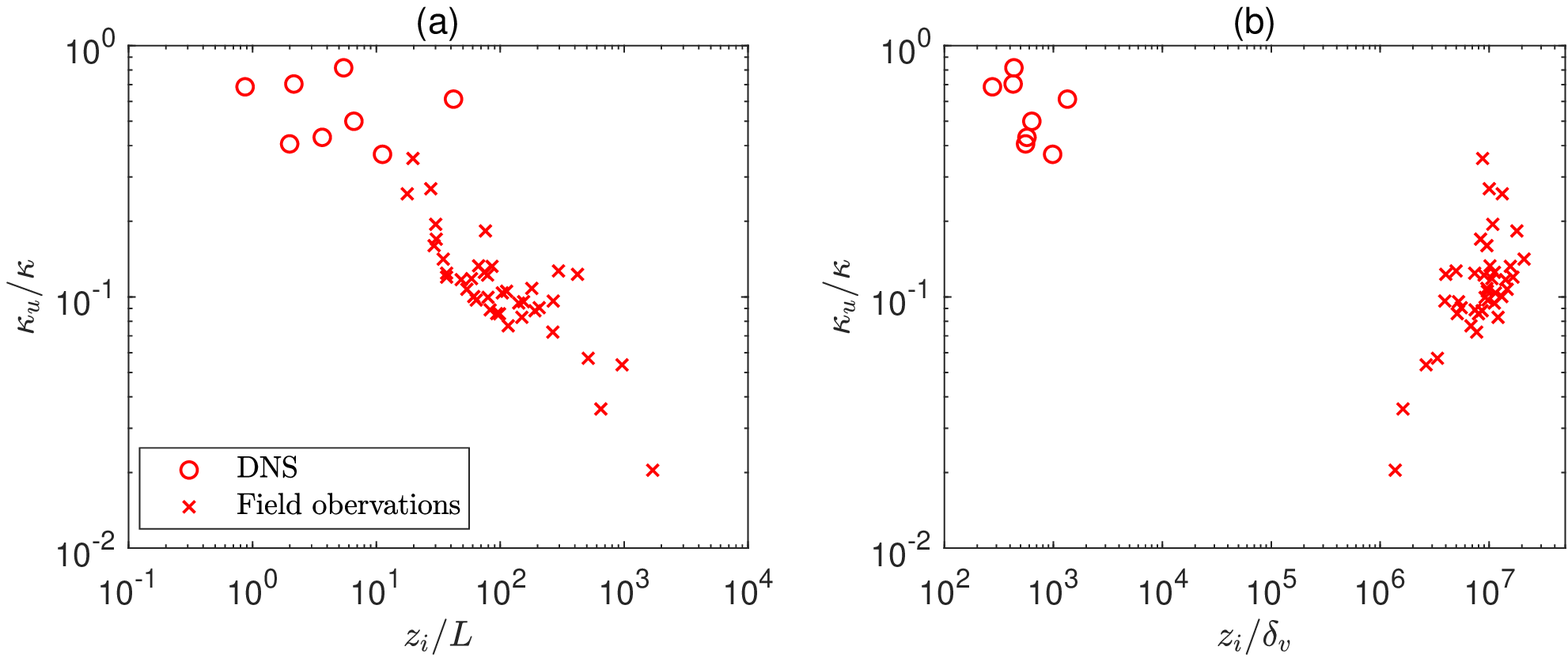}
	\caption{
	The ratio $\frac{\kappa_u}{\kappa}$ plotted against (a) $\frac{z_i}{L}$, and (b) $\frac{z_i}{\delta_v}$ under various stably stratified conditions in DNS experiments and Cabauw observations.} 
	
	\label{fig:DNS_SHEBA_kappau}
\end{figure}

 \subsubsection{DNS and field observations}

The DNS and Cabauw datasets suggest that $\kappa_{u}/{\kappa}$ decreases nonlinearly with increasing $z_i/L$ (Fig. \ref{fig:DNS_SHEBA_kappau}a). That is to say, as buoyancy effects increase (i.e., $z_i/L$ increases), the slope of the stably stratified velocity log law deviates more from that of the neutral channel flow.


When $z_i/\delta_v$ increases from $ \sim 10^3$ (DNS datasets) to $\sim 10^7$ (field observations), $\kappa_u/\kappa$ does not seem to show a monotonic trend (Fig. \ref{fig:DNS_SHEBA_kappau}b).
%
%
%
At high Reynolds number ($1.4 \times 10^6 \le z_i/\delta_v \le 2.1 \times 10^7$), $\kappa_{u}/{\kappa}$ in Cabauw observations can assumed to be independent of Reynolds number thus buoyancy effects dominates. In comparison, the Reynolds number is $276 \le z_i/\delta_v \le 1345$ for the DNS datasets, thus there is a wide separation of Reynolds number between field observations and DNS experiments. The range of MOST stability parameter in the identified log law range are $0.22<z/L<313.70$ and $0.26<z/L<3.67$ for the Cabauw and DNS datasets, respectively. Therefore, both larger Reynolds number and stronger buoyancy effects can be found in Cabauw observations. 
More accurate measurements of turbulent fluxes and denser measurements of velocity in the vertical direction over a wider range of Reynolds numbers and stably stratified conditions as well as laboratory experiments (e.g., Williams et al. \citep{williams2017effect}) will better constrain $\kappa_u$ in the ABL.

\subsubsection{Asymptotic analysis}

 At sufficiently high Reynolds numbers like the field observations, we conduct asymptotic analysis for the slope $\kappa_u$ in the ``neutral limit'' and ``strongly stratified limit'' by neglecting the Reynolds number effects. 

In the neutral limit (where there is no buoyancy) at sufficiently high Reynolds numbers, $z_i/L \rightarrow 0$, we expect that $\kappa_{u}/{\kappa} \rightarrow 1$ since von K\'arm\'an's universal log law is recovered.

In the strongly stratified limit at sufficiently high Reynolds numbers,  $z_i/L \rightarrow \infty$ and $u_\tau \rightarrow 0$, the following asymptotic relation is required to cancel out $u_\tau$:
\begin{equation}
\Phi \bigg( \frac{z_i}{\delta_v}, \frac{z_i}{L} \bigg) =\Phi \bigg(\frac{z_i}{L} \bigg)=\Phi \bigg( \frac{\kappa g z_i}{\Theta_r} \frac{\theta_*}{u_\tau^2} \bigg)=c_1 {\bigg( \frac{\kappa g z_i}{\Theta_r} \frac{\theta_*}{u_\tau^2}  \bigg)}^{1/2} \text{, for }  \frac{z_i}{L} \rightarrow \infty,
\end{equation}
where $c_1$ is a constant. 
Then the velocity gradient ${\partial U}/{\partial z}$ can then be rewritten as
\begin{equation}
\frac{\partial U}{\partial z}= c_1  \left( \frac{\kappa g z_i}{\Theta_r} \right)^{1/2} \frac{\theta_*^{1/2}}{z} \text{, for }  \frac{z_i}{L} \rightarrow \infty.
\end{equation}
The above equation suggests that $\partial U/ \partial z \rightarrow \infty$ as ${z_i}/{L} \rightarrow \infty$ and $\theta_* \rightarrow \infty$, corresponding to extreme stratified conditions.
As $\kappa_u = 1/\Phi$, we can obtain
\begin{equation}
\kappa_u=\frac{1}{c_1} \bigg(\frac{z_i}{L} \bigg)^{-1/2} \text{, for }  \frac{z_i}{L} \rightarrow \infty.
\end{equation}
The slope between $\kappa_u$ and $z_i/L$ obtained from Cabauw observations in the log-log plot is around $-0.4$ (Fig. \ref{fig:DNS_SHEBA_kappau}a), which is not exactly the same as $-1/2$ based on the above equation. However, the asymptotic analysis still qualitatively captures the relation between $\kappa_u$ and $z_i/L$ in extreme stratified conditions at sufficiently high Reynolds number. It is worth noting that the extreme condition ${z_i}/{L} \rightarrow \infty$ is not observed in the Cabauw experiments, which might also be influenced by measurement uncertainties, thus leading to the difference between the observations and asymptotic analysis.

\subsection{Discussion}
The proposed velocity log law in the stably stratified boundary layers can be written as $\frac{\kappa_u z}{u_\tau}  \frac{\partial U}{\partial z}=1$,
or equivalently,
	\begin{equation}  
	\frac{\kappa z}{u_\tau}  \frac{\partial U}{\partial z}=\kappa \Phi \Big( \frac{z_i}{L},\frac{z_i}{\delta_v}\Big) =  \frac{\kappa}{\kappa_u} ,
	\end{equation}
where $\Phi \big( \frac{z_i}{L},\frac{z_i}{\delta_v}\big) $ is independent of $z$ and needs to be determined by numerical experiments or observations.
According to MOST, the normalized velocity gradient was instead assumed to depend on $z/L$ \citep{monin1954basic},
	\begin{equation}   \label{eqnMOST}
	\frac{\kappa z}{u_\tau}  \frac{\partial U}{\partial z}= \phi_m \Big( \frac{z}{L}\Big) ,
	\end{equation}
where $\phi_m$ is a stability correction function dependent on the distance to the wall $z$, thus leading to a non-logarithmic profile. 
The widely used Businger profile \citep{businger1971flux} for MOST is shown in Fig. \ref{fig:DNSloglawKappaU}, which is characterized by a slope rather than the observed plateau for $\frac{\kappa z}{u_\tau}  \frac{\partial U}{\partial z}$.
In numerical experiments, the function $\Phi \big(\frac{z_i}{L},\frac{z_i}{\delta_v} \big)$ does not depend on $z$ thus leading to a log law.
In our various stably stratified DNS datasets,  $\frac{\kappa z}{u_\tau}  \frac{\partial U}{\partial z}$ approaches a constant that is equal to  $\frac{\kappa}{\kappa_u}$ (Fig. \ref{fig:DNSloglawKappaU}), thus supporting a log law rather than MOST.
In addition, the slope of the proposed velocity log law depends on $\frac{z_i}{L}$ (buoyancy effects) and $\frac{z_i}{\delta_v}$ (Reynolds number effects). Such dependence of the slope on $\frac{z_i}{L}$ and $\frac{z_i}{\delta_v}$ has also been reported in temperature profiles in the convective boundary layers \citep{cheng2021logarithmic}.

  \section{Conclusion}
  
 We report new logarithmic velocity profiles in the near-wall region affected by buoyancy effects, through dimensional analysis, DNS experiments, and field observations of the stably stratified boundary layers. The new velocity log law can be described by $\frac{\kappa_u z}{u_\tau}  \frac{\partial U}{\partial z}=1$, where $\kappa_u$ is a function of $\frac{z_i}{L}$ (buoyancy effects) and $\frac{z_i}{\delta_v}$ (Reynolds number effects) supported by both the DNS experiments and field observations. 
Asymptotic analysis for the slope $\kappa_u$ has been conducted in the neutral limit and strongly stratified limit at sufficiently high Reynolds numbers.
%
More accurate observations over a wider range of Reynolds numbers and stably stratified conditions may better constrain $\kappa_u$ in the atmosphere. 
The proposed velocity log profile may replace Monin-Obukhov similarity function for velocity in global climate models and wall models for large eddy simulations, possibly leading to more realistic predictions of weather, climate, and hydrology, especially in polar regions.

\begin{acknowledgments}
We would like to thank Dr. Kaighin McColl for helpful discussions. The computations in this paper were run on the FASRC Cannon cluster supported by the FAS Division of Science Research Computing Group at Harvard University. We would like to thank Dr. Fred C. Bosveld and Henk Klein Baltink for the help in obtaining the field data at the Cabauw Experimental Site for Atmospheric Research (http://www.cesar-database.nl).
\end{acknowledgments}

\appendix




%

\end{document}